\documentclass[pdflatex,sn-mathphys-num]{sn-jnl}

\usepackage{graphicx}%
\usepackage{multirow}%
\usepackage{amsmath,amssymb,amsfonts}%
\usepackage{amsthm}%
\usepackage{mathrsfs}%
\usepackage[title]{appendix}%
\usepackage{textcomp}%
\usepackage{manyfoot}%
\usepackage{booktabs}%
\usepackage{algorithm}%
\usepackage{algorithmicx}%
\usepackage{algpseudocode}%
\usepackage{listings}%

\usepackage{physics} 
\usepackage{tikz}
\usetikzlibrary{decorations.pathreplacing}
\usetikzlibrary{automata, positioning, arrows, calc,math,arrows.meta}

\theoremstyle{thmstyleone}%
\theoremstyle{thmstyletwo}%

\theoremstyle{thmstylethree}%
\newtheorem{definition}{Definition}%

\begin{document}

\title[Implementing a Quantum Finite Automaton in IBMQ using Custom Control Pulses]{Implementing a Quantum Finite Automaton in IBMQ using Custom Control Pulses}


\author*[1]{\fnm{Eduardo} \sur{Willwock Lussi}}\email{eduardolussi@gmail.com}

\author[1]{\fnm{Lucas} \sur{Cavalcante de Sousa}}\email{lucks.sousa@gmail.com}
\equalcont{These authors contributed equally to this work.}

\author[1]{\fnm{Jerusa} \sur{Marchi}}\email{jerusa.marchi@ufsc.br}
\equalcont{These authors contributed equally to this work.}

\author[1]{\fnm{Rafael} \sur{de Santiago}}\email{r.santiago@ufsc.br}
\equalcont{These authors contributed equally to this work.}

\author[2]{\fnm{Eduardo} \sur{Inacio Duzzioni}}\email{duzzioni@gmail.com}
\equalcont{These authors contributed equally to this work.}

\affil*[1]{\orgdiv{Department of Informatics and Statistics}, \orgname{Federal University of Santa Catarina}, \orgaddress{\street{Delfino Conti St.}, \city{Florianópolis}, \postcode{88.040-370}, \state{Santa Catarina}, \country{Brazil}}}

\affil[2]{\orgdiv{Department of Physics}, \orgname{Federal University of Santa Catarina}, \orgaddress{\street{Desembargador Vitor Lima St.}, \city{Florianópolis}, \postcode{88.040-900}, \state{Santa Catarina}, \country{Brazil}}}


\abstract{Quantum finite automata can be used for pattern recognition. Present implementations on actual quantum devices face decoherence issues, which compromise the quality of long strings computation. In this work, we focus on Measure Once 1-way Quantum Finite Automata (MO1QFA) model for addressing the $\text{MOD}^p$ problem, investigating how quantum errors may affect the quality of the computation in this model when implemented in IBM-Q superconducting environment. To improve the performance of the implementation, we use pulse-level programming for calibrating a fast single-qubit gate designed specifically for the automaton implementation. The demonstrations conducted on Jakarta quantum computer show that using custom pulses significantly reduces errors during extended word computations. While realizing improvements in error variations and predictability — with a fourfold reduction in circuit latency - the proposed solution demonstrates a substantial increase in the supported computation length of the automaton. When considering thresholds of 10\% and 20\% in absolute errors of acceptance probabilities, the solution has the potential to increase the maximum word length by 12 and 7$+$ times, respectively, compared to the default Qiskit gate.}


\maketitle

Finite Automaton (FA) is the most elementary classical computing model. Despite their simplicity, finite automata find practical application in various domains in classical computing, including tasks such as lexical analysis and text search \cite{sipse2012introduction}.

Quantum computing brings forth a distinctive perspective on complexity, rendering solutions to classical difficult problems with possible exponential speedups. In the realm of automata theory, the advent of Quantum Finite Automata (QFA) heralds a heightened potential in formal languages. QFA, in particular, exhibit the capacity to embrace certain Context-Free Languages while employing considerably smaller memory resources \cite{MM1QFA}.

Several endeavors have been made to implement Quantum Finite Automata (QFA) in quantum hardware. A photonic quantum automaton implementation was undertaken \cite{PhotonicQFA} and subsequently refined \cite{EnhancedPhotonicQFA}. However, utilizing the polarization degree of freedom introduces challenges in recognizing extensive strings because it requires the insertion of a rotator for each word. In superconducting technologies, an implementation of a QFA for the $\text{MOD}^p$ problem was demonstrated within the IBMQ platform \cite{IBMQ}, but default implementation suffered from decoherence \cite{QFANoisy}.

In fact, despite significant advancements, even more after demonstrations of quantum advantage \cite{QuantumSupremacy,wu2021strong,zhong2021phase,zhu2022quantum,zhong2020quantum,madsen2022quantum}, the fragility of quantum systems persists as a prominent challenge. Present-day qubit implementations remain markedly susceptible to environmental noise, nuances in qubit control techniques, variations in measurement equipment, and other stochastic error processes that culminate in decoherence \cite{KrantzQuantumEngineersGuide}. This phenomenon results in the degradation of quantum behavior, effectively reducing the system's characteristics to a classical state \cite{OutlookQC}, consequently eroding both the reliability and efficiency of quantum computations.

There are mainly two approaches for improving the execution performance of quantum algorithms. The first, quantum error correction, acts in a higher level of abstraction, by reducing the error probability through the redundant encoding of logical qubits into a larger number of physical qubits \cite{OutlookQC}. The second works in a lower level of abstraction. Quantum control can improve the performance of quantum hardware by efficiently manipulating the qubits for having the desired behavior \cite{QCTRL_QuantumControl}. In superconducting quantum computing, quantum gates are translated into microwave pulses that are capable of changing the state of the quantum system. Minor oscillations of this control or bad control pulses may cause decoherence. In IBMQ, superconducting quantum computers can be physically controlled by Qiskit Pulse \cite{QiskitPulseDoc, QiskitPulse}, which is a pulse-level programming kit.

Recent research has underscored the potential of leveraging quantum control to enhance both the fidelity of quantum gates and the overall hardware control, resulting in improved computational outcomes \cite{ErrorRobustAndre, DRL_QCTRL, Leakage, Bosonic, microarchitecture, Learning, Propson2022, Mundada2023}. This advancement has also given rise to commercial solutions in the field, exemplified by Q-CTRL \cite{QCTRL_site}, a company that has introduced two products specifically geared towards error mitigation through quantum control techniques \cite{ErrorRobustAndre, DRL_QCTRL, Mundada2023}. The initial offering \cite{ErrorRobustAndre} provides an interface designed to engineer noise-robust quantum gates. This innovation draws from optimization strategies rooted in the thorough characterization of error processes. On the other hand, the second product \cite{DRL_QCTRL} introduces an autonomous method for gate generation and hardware calibration that circumvents the need for prior knowledge about the system, its control mechanisms, and error processes. The solution was integrated into an automated workflow featuring additional compilation optimizations in \cite{Mundada2023}, showcasing improvements exceeding 1000x in success probabilities for quantum algorithms on the IBMQ platform. The complexity and elusiveness of error processes make their full comprehension and characterization challenging. In response, Machine Learning (ML) techniques have emerged as potent tools for error mitigation, capable of alleviating the requirement for exhaustive error process characterization \cite{DRL_QCTRL, Nautrup2019optimizingquantum, Learning}. The efficacy of ML-driven approaches remains reliable even when confronted with intricate error dynamics, as illustrated in \cite{DRL_QCTRL, Nautrup2019optimizingquantum, Learning, Mundada2023}.

Within this paper, we carry out the Measure Once 1-way Quantum Finite Automaton (MO1QFA) \cite{moore2000quantum} in the IBM Quantum platform \cite{IBMQE}. Our investigation delves into the performance of MO1QFA within this platform and extends to the impact of errors on lengthy string computations. Motivated by challenges identified in \cite{QFANoisy}, we further introduce the application of pulse-level programming techniques. These techniques facilitate the design and calibration of custom quantum gates tailored specifically for the automaton's implementation. Through this avenue, we significantly surpass the performance of existent single-qubit implementations and default Qiskit quantum gates.

This paper is organized as follow. The definition of QFA is presented in Section \ref{sec:finite_automata}. An introduction about the Quantum Automaton Model and the modeled language under our consideration is given in Section \ref{sec:mo1qfa}. Section \ref{sec:pulse_level_programming} briefly presents the pulse-level programming in superconducting quantum computers. After, Section \ref{sec:experiments} shows the demonstrations conducted and the obtained results with the MO1QFA implementation. Finally, Section \ref{sec:conclusion} summarizes the results and gives some directions for future works.

\section{Finite Automata}
\label{sec:finite_automata}

Classical finite automata are a simple model of computation used to solve many theoretical and practical problems in classical computing. In formal language theory, finite automata are able to recognize the class of Regular Languages. Also, they can be deterministic or non-deterministic, but both representations are equivalent \cite{sipse2012introduction}.

\begin{definition} A Non-Deterministic Finite Automaton is defined as a 5-tupla $M = (Q, \Sigma, \delta, q_0, F)$ where:
\begin{itemize}
\item $Q = \{q_0, q_1, \cdots, q_n\}$ is the set of finite states; 
\item $\Sigma = \{\alpha_1, \alpha_2, \cdots, \alpha_m\}$ is the finite set of input symbols (alphabet);  
\item $\delta$ is the transition function defined as $\delta: Q \times \Sigma \rightarrow 2^Q $, where $2^Q$ is the power set of $Q$; 
\item $q_0 \in Q$ is the initial state; and 
\item $F \subseteq Q$ is the set of final states.    
\end{itemize}
\end{definition}

The computation process starts with a string $W$ recorded on the machine tape and the machine reading head positioned in the most left character of $W$. At each step, according its internal state $q$ and the symbol $\alpha$ over the reading head, the machine applies the transition function $\delta(q, \alpha)$, changing its internal state and moving the reading head one cell to the right. This computing process finishes when all the string $W$ is read. The string is classified as belonging to the machine language \footnote{A language is a decision problem and strings in a language are instances of this problem.} if a final state is reached, that is, the extended transition $\widehat{\delta}(q_0,W) \rightarrow q_i$ with $q_i \in F$. 

\section{Measure Once 1-Way Quantum Finite Automaton}
\label{sec:mo1qfa}

Quantum finite automata are the quantum counterparts of the classical ones. Because of their probabilistic nature, they can be much more powerful, recognizing some Context Free Languages. These quantum models can be classified by their measurements, reading head movement, transitions and some other features, that is, there are measure-many \cite{MM1QFA} and measure-once \cite{moore2000quantum}, 1-way \cite{MM1QFA, moore2000quantum}, 2-way \cite{MM1QFA, 2CQFA} or either 1.5-way \cite{15QFA} quantum finite automata described by unitary \cite{moore2000quantum} or non-unitary transformations, and with many peculiar characteristics as the use of some classical automata features \cite{2CQFA}.

The Measure-Once 1-way Quantum Finite Automaton (MO1QFA) was proposed in 2000 by Moore and Crutchfield \cite{moore2000quantum}. Its main characteristics are: 

\begin{itemize}
    \item State transitions are performed as unitary transformations;
    \item The quantum state is measured once at the end of the execution;
    \item Reading head movement is one-way, i.e., it moves from left to right reading a symbol at once after each transition;
    \item At the beginning of computation, the reading head is at the first symbol of the tape.
\end{itemize}

Because of these restrictions MO1QFA is considered the simplest quantum finite automaton model and also the less expressive model when compared with other quantum finite automata models \cite{MM1QFA}. It is defined as follows:\\

\begin{definition}
    MO1QFA is a 5-tupla $M = (Q,\Sigma,\delta,\ket{q_0},F)$ where:
    \begin{itemize}
        \item $Q = \{\ket{q_0}, \ket{q_1}, \hdots, \ket{q_n}\}$ is the finite set of states - $Q \subseteq \mathcal{H}$, where $\mathcal{H}$ is the Hilbert space;
        \item $\Sigma = \{\alpha_0, \alpha_1, \hdots, \alpha_m\}$ is the finite set of input symbols (alphabet);
          \item $\delta$ is the set of unitary matrices $U_{\alpha_i}$, where each $U_{\alpha_i}$ describes the state transition according to $\alpha_i$;
        \item $\ket{q_0}$ $\in Q$ is the automaton initial state;
        \item $F$ $\subseteq Q$ is the set of final states.
    \end{itemize}
    
    The computing process over a string $W = w_0w_1w_2\ldots w_{\mid W\mid -1} $, where $w_i \in \Sigma$ in MO1QFA is given by
    \begin{equation}
        f(W) = \norm{P_{acc}U_W\ket{q_0}}^2,
    \end{equation}
    in which $U_W$ means apply the respective unitary matrix to each symbol $w_i$ of $W$
    \begin{equation}
        U_W = U_{w_{\mid W \mid-1}} \hdots U_{w_1}U_{w_0},
    \end{equation}
    $P_{acc}$ is the automaton final states projection operator
    \begin{equation}
        P_{acc} = \sum_{q_i\in F} \ketbra{q_i}{q_i},
    \end{equation}

\noindent and $f(W)$ represents the probability of the automaton $M$ to accept $W$.
\end{definition}

\subsection{\label{sec:languages}The $MOD^p$ Language}

\textit{Say} and \textit{Yakary{\i}lmaz} \cite{say2014quantum} proposed to use MO1QFA to solve problems that have limited error using fewer states when compared to the same solution in classical finite automata. For demonstrating this, they studied the prime number modulus problem:

\begin{equation}
    \text{MOD}^p = \{a^{jp} \mid j \text{ is a non-negative integer}\}.
\end{equation}

\noindent In this language, $a$ is the input symbol and strings belonging to this language have a length congruent to $0$ modulo $p$, that is $\abs{W} \mod p = 0$.

In classical automata, the solution for this problem consists of $p$ states. The initial state represents $\text{MOD}^p=0$, which also serves as the final state. Subsequent states represent $\text{MOD}^p=1$ through $\text{MOD}^p=p-1$. When reading input symbol $a$, a transition occurs from state $\text{MOD}^p=i$ to state $\text{MOD}^p=i+1$, except when transitioning from state $\text{MOD}^p=p-1$ to $\text{MOD}^p=0$.

On the other hand, it has been shown that the number of states, which is the complexity measure for finite automata, in 1-way QFAs is exponentially smaller than classical automata \cite{ambainis19981}. This specific language can be modeled with only two states, the qubit basis states $\ket{0}$ and $\ket{1}$. The transition matrix describes rotations on the $x$ axis by an angle that depends on the prime number $p$ in the following form:

\begin{equation}
  U_a(p) =
  \begin{bmatrix}
    \cos(\frac{2\pi}{p}) & -i\sin(\frac{2\pi}{p})\\
    -i\sin(\frac{2\pi}{p}) & \cos(\frac{2\pi}{p})
  \end{bmatrix}.
  \label{mod:matriz}
\end{equation}

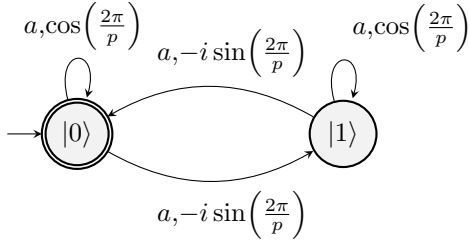
\begin{figure}[h]
  \centering
  \begin{tikzpicture}[->, >=stealth, node distance=3.5cm, every state/.style={thick, fill=gray!10},initial text=$ $]
    \node[state, initial, accepting] (0) {$\ket{0}$};
    \node[state, right of=0] (1) {$\ket{1}$};
    \draw
    (0) edge[loop above] node[text width=1.4cm,align=left]
    {$a$,$\cos(\frac{2\pi}{p})$} (0)
    (0) edge[bend right, below] node[text width=1.4cm,align=center]
    {$a$,$-i\sin(\frac{2\pi}{p})$} (1)
    (1) edge[loop above, above right] node[text width=1.4cm,align=center]
    {$a$,$\cos(\frac{2\pi}{p})$} (1)
    (1) edge[bend right, above] node[text width=1.4cm,align=center]
    {$a$,$-i\sin(\frac{2\pi}{p})$} (0);
  \end{tikzpicture}
  \caption{MO1QFA automaton diagram for $MOD^p$ problem. $\ket{0}$ is the initial state (indicated by the horizontal arrow) and also the final one (represented in the double circle), and $\ket{1}$ is an intermediate state of the computation process. After reading a symbol $a$ of the string, the result of the computation will depend on the amplitudes of remaining in the same state $\cos(2\pi/p)$ or changing to another state $-i\sin(2\pi/p)$, which are described above or under the arrows.}
  \label{modp:diagrama}
\end{figure}

The action of matrix $U_a(p)$ and the whole performance of the automaton can be visualized in Fig. \ref{modp:diagrama}. The initial and final state is $\ket{0}$, represented by the double circle, so that the probability of acceptance of a word with length $w$ is $\cos^2(2\pi w/p)$. We observe that after a given number $jp$ of applications, with $j \in \mathbb{Z}_+$, the probability of acceptance is $1$, then the string is said to be accepted.

Although the automaton accurately recognize strings of length congruent to $0$ modulo $p$ with $100\%$ probability, it is prone to false positives. The upper limit of the error can be observed in string lengths that are close to being congruent to $p/2$ modulo $p$ \cite{say2014quantum}. In such cases, the error approximates $\cos^2\left(\pi/p\right)$. Furthermore, as $p$ increases and string length approaches values congruent to $p/2$ modulo $p$, the error rate rapidly escalates \cite{say2014quantum}, making these values to be mistakenly accepted by the automaton with more than 80\% of probability for primes greater than 7.

To achieve a fixed error bound, however, we can execute multiple 2-state MO1QFA in parallel using different rotation angles in the form $k\times2\pi/p$, where $k \in \left\{1, \cdots, p-1\right\}$. Previous work has demonstrated that this can be accomplished with the appropriate set of $k$ values \cite{ambainis19981}.

\section{Pulse-level Programming in Quantum Computing}
\label{sec:pulse_level_programming}

IBMQ superconducting quantum computers can be physically controlled by Qiskit Pulse \cite{QiskitPulseDoc, QiskitPulse}, which is a pulse-level programming kit that allows to extract greater performance from quantum hardware. Specifically in superconducting quantum computers, the gate-level translation to pulse-level is done by microwave pulses, which interact with the qubits changing their quantum state.

Pulse-level programming is a complicated task. Basic descriptions of pulses are composed of three variables: frequency, amplitude, and time. Frequency corresponds to the value in the microwave spectrum which causes transitions between energy levels \cite{SuperconductingPennyLane} (e.g. $\ket{0} \leftrightarrow \ket{1}$). Although IBMQ backends provide reasonably accurate estimates of qubit frequencies, the available documentation \cite{CalibratingQiskit} demonstrates an experimental determination of qubit frequencies, which can yield better results because this value may fluctuate over time. The strength of a control pulse is determined by its amplitude. The effect on the qubit state is observed through the Rabi experiment, which consists of testing quantum hardware to evaluate its behavior for specific pulses. When a pulse is applied, it causes the quantum state to undergo an oscillation. This oscillation varies depending on the pulse amplitude: the higher the amplitude, the faster the quantum state oscillates. Therefore, the Rabi experiment establishes the relationship between the amplitude and time of a pulse and the rotation of the state vector (known as Rabi oscillation) in a quantum state. The Rabi experiment results allow us to discover the Rabi frequencies, which are the frequencies at which the state vector oscillates along a specific axis. Knowing the Rabi frequencies enables the design of different pulses that have the same effect in the quantum state, i.e., longer pulses with smaller amplitudes or faster pulses with larger amplitudes.

Since different pulses can have the same effect in the quantum state, designing quantum gates can be very difficult. Shorter pulses may have a better performance because the circuit duration holds off decoherence, but shorter pulses require greater amplitudes, which can cause leakage errors \cite{LeakageFastPulse}. The durability of pulses is also a problem, the qubit frequency changes over time and it is necessary to periodically recalibrate quantum gates and qubit's frequency. Additionally, the pulse design needs to consider decoherence and other control errors. Decoherence is frequently measured by $T_1$ and $T_2$ times. $T_1$ is the time taken for the quantum state $\ket{1}$ to decay to the quantum state $\ket{0}$ and $T_2$ is the dephasing time, that is, the time in which the quantum system can preserve its relative phase information of the quantum state \cite{QC_ImplementationIssues}. Besides, control errors are related to the quantum control of the quantum system, such as amplitude control errors \cite{ErrorRobustAndre}, leakage \cite{Chen2016leakage}, crosstalk \cite{Sarovar2020detectingcrosstalk} and errors in the measurement of the quantum state \cite{Funcke2022}.

In the context of Qiskit and IBMQ quantum computers, specific prerequisites govern the construction of pulse schedules. A fundamental building block is the sample time ($dt$), which serves as the foundational temporal unit for pulse scheduling \cite{QiskitDocSystemInfo}. Every pulse schedule is defined in relation to this sample time. Some backends impose restrictions on pulse length. In the quantum computer used in this work, Jakarta, the shortest pulse schedule is $80dt$. However, other backends may require longer schedules \cite{QiskitDocBuildSchedules}. Additionally, the pulse duration must be a multiple of 16 in this backend. Various methodologies exist for describing a pulse \cite{QiskitPulseDoc}. The simplest approach involves utilizing \textit{waveform}, which is essentially a list of complex amplitude values, each of these values corresponds to an application in the $dt$ time. Alternatively, parametric pulses can be employed to depict different shapes, such as square, Gaussian, Gaussian Square or the default shape prevalent in most backends—the Derivative Removal by Adiabatic Gate (DRAG) \cite{QiskitDocDRAG042}.

Nevertheless, most quantum algorithms implementations are limited to the quantum system resources in terms of computation quality. IBMQ systems maintain a determined universal gate set which, individually, is very robust. However the need to transpile the original circuit to the available gate set excessively increases the circuit duration mostly, which then, compromises the computation quality. This arrangement imposes the design of personalized quantum gates to improve performance. On the subject of quantum automata, this paper offers an implementation using custom quantum gates that can considerably improve the performance in the computation of long strings.

\section{Demonstrations and Outcomes}\label{sec:experiments}

To demonstrate the implementation of a quantum automaton in a real-world environment, we consider a MO1QFA designed to solve the $\text{MOD}^p$ problem, introduced in section \ref{sec:languages} for the case of $p=11$. For the quantum environment, we utilize IBMQ \cite{IBMQ} to assess its suitability as a platform for quantum automaton implementation, taking into account quantum errors. The experimental evaluations were carried out using qubit 0 of Jakarta quantum computer. Each demonstration was replicated four times on different days. For each circuit, a total of 1024 shots were executed.

Regarding the MO1QFA implementation, our investigation covers word lengths that are congruent to both 0 and 3 modulo 11, within the range from 0 to 1000, resulting in a total of 182 distinct values. With these values, the expected acceptance probability is 100\% for word lengths congruent to 0 modulo 11, and 2.0254\% for word lengths congruent to 3 modulo 11. 

For implementing the automaton, we leverage the hardware of Jakarta for calibrating a fast square pulse, which in turn, brings about an improved gate performance. The demonstrations conducted within this section shed light on the extent to which errors can impact quantum automata implementations. In order to grasp the varying roles played by distinct techniques in optimizing our gate, a comparative analysis is carried out, contrasting the default Qiskit gate implementation devoid of any circuit optimizations with the calibrated gate.

\subsection{MO1QFA Circuit Mapping}

The MO1QFA model has only two states, consequently, the circuit implementation requires only one qubit. The transitions can be carried out using a $R_x(\theta)$ gate, which performs a $\theta$ rotation around the $x$ axis. So the mapping from MO1QFA (see Eq. (\ref{mod:matriz})) to a Qiskit circuit model is given by a sequence of $R_x(4\pi/11)$ gates that describe the rotations applied by the transitions taken during the execution of an input in the automaton.

To optimize the circuits that run at IBMQ backends, Qiskit implements four levels of optimization, 0 through 3, the higher the number, the more optimized it is, at the expense of more time to transpile \cite{QiskitPassManager}. Optimization level 0 just maps the circuit to the backend, making no optimizations. Optimization level 1 makes some lightweight optimizations by collapsing adjacent gates (default). Optimization levels 2 and 3 make medium and heavy-weight optimizations considering gate communications relationships. In our case, as we are only using single-qubit gates, the optimization levels 2 and 3 are not relevant.

For implementing the automaton on IBMQ, optimization level 1 produces the best results. This occurs because this optimization combines adjacent gates. Instead of applying each of the $R_x(4\pi/11)$ gates individually, it applies only one $R_x((|W| \; \text{mod} \; 11) \times 4\pi/11)$ gate. Although this approach does yield the best results, it is not accurate for the implementation, as the concept assumes that the automaton itself performs the computation, with the length of the string unknown to it. As a result, when using this optimization, it is the classical computer executing the optimization that performs the computation, not the quantum automaton.

The situation is similar when transferring the rotation to the $z$ axis, using $R_z$ rotations instead of $R_x$, which also produces satisfactory results. However, in this instance, the effectiveness is due to the fact that $z$ rotations are not directly implemented in IBMQ quantum computers. $R_z$ gates are interpreted as a phase shift on subsequent pulses \cite{McKay2017}.

Thus, optimization level 0 should correctly implement the automaton, applying the single-qubit gate $R_x(4\pi/11)$ every time a symbol in the string $W$ is read. Consequently, due to decoherence and quantum gate errors, the longer the word, it is expected that there will be an increasing error in the state of the qubit at the end of the computation.

\subsection{Qiskit Default Gate Implementation Performance}

\begin{figure}[h]
\centering
\includegraphics[width=0.8\textwidth]{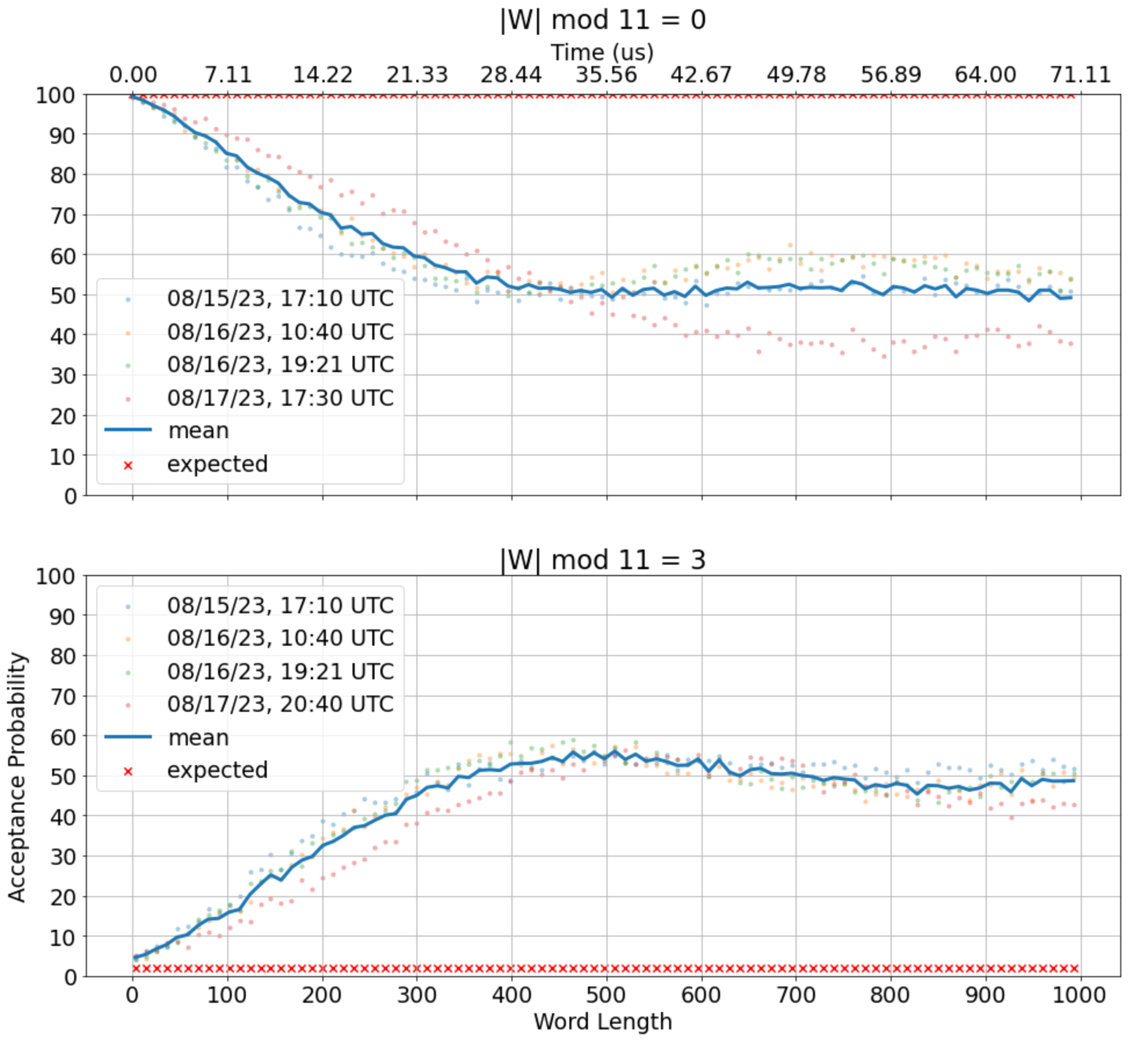}
\caption{Comparison between expected probability for word lengths congruent to 0 and 3 modulo 11 and acceptance probability with optimization level 0 run on \textit{ibm\_jakarta}.}
\label{fig:2x160dt_default-drag}
\end{figure}

The first demonstrations use the default Qiskit gate implementation with no circuit optimizations to perform the quantum automaton. Fig. \ref{fig:2x160dt_default-drag} shows a comparison between the expected (crosses) and obtained (dots and line) acceptance probabilities from Jakarta for word lengths congruent to 0 and 3 mod 11 with the word length ranging from 0 to 1000.

As remarked previously, there is an error accumulation as the word length increases. The minimum error happens when the word is at its minimum length, and as it increases, the error also increases until it becomes arbitrary. For both word categories, Jakarta presented crescent errors until around 500 pulse repetitions, where absolute errors between the obtained and expected were around 50\%.

\begin{figure}[h]
\centering
\includegraphics[width=0.6\textwidth]{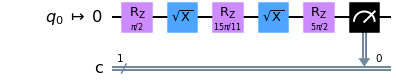}
\caption{Transpiled quantum circuit for $R_x(4\pi/11)$ gate.}
\label{fig:rx_circuit}
\end{figure}

The primary factor contributing to the errors in this implementation is the quantum computer gate set dependence. During the transpilation process with optimization level 0, the high-level quantum gates, such as the $R_x$ gates in this scenario, need to be mapped onto the available quantum gates of the quantum computer. To illustrate, Figure \ref{fig:rx_circuit} demonstrates the realization of the $R_x(4\pi/11)$ gate within this quantum system. It is implemented as a sequence of five other quantum gates, significantly elongating the circuit and compromising the algorithm's performance. Notably, this particular sequence holds interest due to the fact that each single-qubit gate can be constructed by calibrating a single $\sqrt{X}$ gate. Moreover, the $R_z$ gates are virtually implemented by adjusting the phase of subsequent pulses \cite{McKay2017}. Despite these efforts, this sequence still consumes a duration of $320dt=71.11$ $ns$. Figure \ref{fig:2x160dt_default-drag} also shows a timescale of the circuit execution, highlighting that word lengths near 1000 necessitate almost $71.11$ $\mu s$ – a duration that exceeds the $T_2$ time of approximately $45$ $\mu s$ (on average) and approaches the $T_1$ time of around $146$ $\mu s$ (on average). Additional information about backend configuration can be found in Appendix \ref{apx:jobs-info}.

Furthermore, IBMQ employs the Derivative Removal by Adiabatic Gate (DRAG) for the $\sqrt{X}$ gate. This pulse shape is fundamentally a Gaussian waveform augmented with an extra Gaussian derivative and lifting components. The usefulness of this gate lies in its resilience to leakage errors \cite{Gambetta2009, Gambetta2011}, when energy levels beyond $\ket{1}$ are excited. While it effectively accomplishes this objective, it is hard to achieve efficient fast DRAG pulses due to the need for high amplitude values. This limitation stems from the inherent Gaussian nature of the pulse, resulting in the loss of its amplitude values half backward and half forward from the maximum amplitude value in the middle, in comparison to, for instance, a square pulse, which maintains a constant amplitude and the lowest amplitude value.

These demonstrations reveal that the automaton does not perform well using the default gate implementation. The strategy followed by default in IBMQ for single-qubit gate implementation aims to achieve any unitary operation while minimizing gate calibration needs. Therefore, when implementing a specific gate like the $R_x$ gate used by the automaton, errors arise from both the generalization of the quantum gate implementation and the decoherence induced by the increased pulse length. Additionally, the repetitive nature of automata amplifies these errors. But QFAs have well-defined sets of operations, typically small enough to be calibrated accurately. Consequently, performance improvements are expected when using a custom calibrated gateset, as it allows for the design of faster and more specific control pulses that are less affected by error accumulation and decoherence during the repetitive computation of automata.

\subsection{Fast Square Gate Performance}\label{sec:fast-square-gate-performance}

Given the challenges associated with the default implementation of the $R_x$ gate, which decomposes into two $\sqrt{X}$ gates with phase shifts, executed through DRAG pulses, the demonstrations detailed in this section center around the creation of a fast square $R_x(4\pi/11)$ gate. This pursuit aims to enhance the efficiency of computational processes for longer sequences. The integration of fast pulses holds the potential to curtail circuit duration and distance it from the quantum computer's decoherence time. To achieve this goal, we followed Q-CTRL Boulder Opal documentation elaborated in \cite{QCTRLCalibration} and \cite{QCTRLDesignGates}. These resources help in designing the gate and precisely calibrating the hardware, with the goal of improving the outcomes obtained from previous pulse repetitions.

In the scope of this research, the following Hamiltonian is used to describe the evolution of the quantum system
\[
  H\left(t\right)=\frac{1}{2}\left(\gamma^*\left(t\right)\sigma_- + \gamma\left(t\right)\sigma_+\right)=\frac{1}{2}\left(I\left(t\right)\sigma_x + Q\left(t\right)\sigma_y\right),
\]
where $\gamma\left(t\right)=I\left(t\right) + iQ\left(t\right)$ is the time-dependent complex-valued control pulse waveform described in terms of Rabi frequencies and $\sigma_k$, with $k=x, y, z$ are the Pauli matrices.

The waveform of the first pulse used here is a square pulse with duration $80$ $dt=17.8$ $ns$, as fast as possible for this backend, whose imaginary amplitude ($A_Q$) equals $0$ and the real part ($A_I$) is the amplitude value needed for a Rabi frequency about $I=2\pi \times 10.23$ $MHz$, which is the value required for a $4\pi/11$ rotation around the $x$ axis in the Bloch sphere in $17.8$ $ns$. The control pulse waveform described needs to be calibrated based on the target quantum hardware to work properly. To do this, it is necessary to perform the Rabi experiment identifying the relationship $\left(A_I, A_Q\right) \leftrightarrow \left(I, Q\right)$, that is, find the pulse amplitudes $A_I$ and $A_Q$ related to the Rabi frequencies $I\left(t\right)$ and $Q\left(t\right)$, respectively, in the Hamiltonian. 

\begin{figure}[h]
\centering
\includegraphics[width=0.7\textwidth]{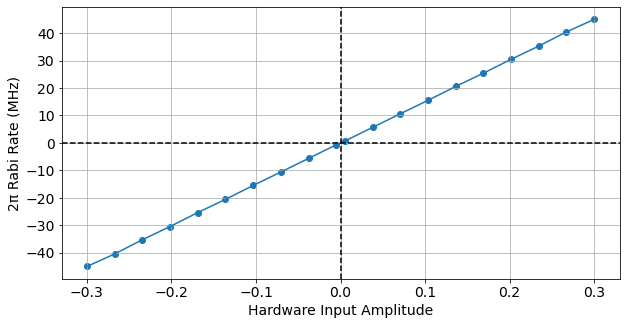}
\caption{$2\pi$ Rabi rates calibration for Jakarta.}
\label{fig:rabi-rates}
\end{figure}

Figure \ref{fig:rabi-rates} illustrates the Rabi rate calibration for $2\pi$ rotations, acquired through an interpolation over 10 amplitude values, selected from the range of 0.005 to 0.3. The results are mirrored for negative values. The initial experiment involves observing the evolution of square pulses with varying amplitude values – in this instance, 10 values are examined. For each experiment, a sequence of measurements is performed, and the observed evolution is fitted to the following function:
\[
    y = A \times cos^2\left(2\pi \times \textit{rabi\_rate} \times x + \phi \right),
\]
where $A$ and $\phi$ denote tuning parameters, $x$ is time, $y$ represents the projection onto the $z$ axis of the quantum state, and \textit{rabi\_rate} signifies the Rabi rates depicted in Figure \ref{fig:rabi-rates} for $2\pi$ rotations. Hence, from the Rabi calibration plot, we can deduce the appropriate amplitude values corresponding to each Rabi rate. In this instance, a Rabi frequency of approximately $2\pi \times 10.23$ $MHz$ necessitates an amplitude of around 0.068.

\begin{figure}[h]
\centering
\includegraphics[width=0.8\textwidth]{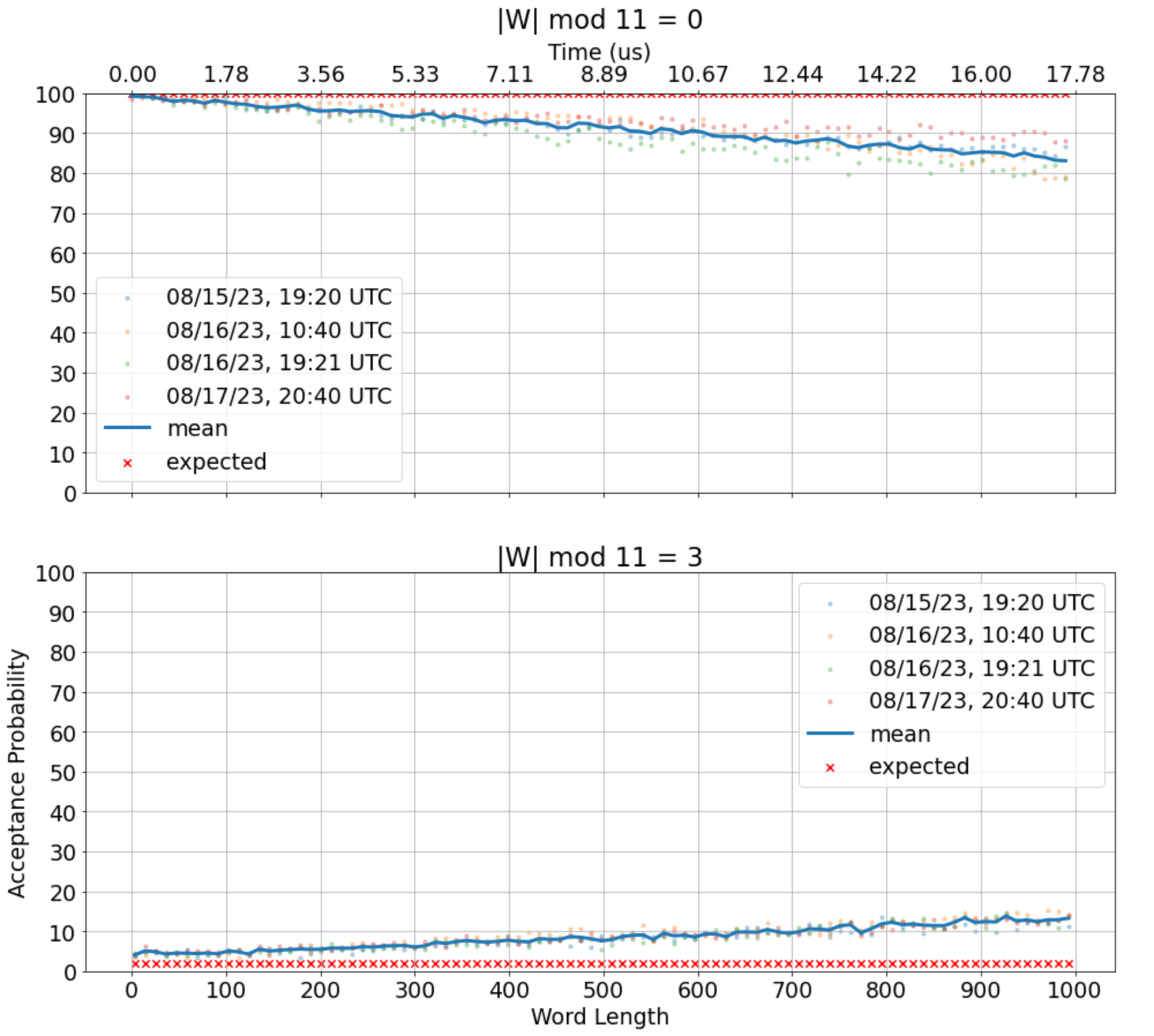}
\caption{Expected and obtained acceptance probabilities for the $17.8$ $ns$ square pulse for word lengths congruent to 0 and 3 modulo 11.}
\label{fig:80dt_square}
\end{figure}

For the automaton implementation, the simple and fast square pulse generation using basic Rabi calibration was able to considerably improve the results. Figure \ref{fig:80dt_square} shows the results for word lengths congruent to $0$ and $3$ modulo $11$, respectively, implemented on Jakarta quantum computer. These results are very intriguing in comparison to the original $R_x (4\pi/11)$ implementation (Fig. \ref{fig:2x160dt_default-drag}), since the acceptance probabilities deviations increase much slower.

When contrasted with the default implementation of the $R_x(4\pi/11)$ gate on the Jakarta platform, the adoption of a shortened pulse duration has demonstrated remarkable efficiency. In the demonstrations involving the default approach, as illustrated in Figure \ref{fig:2x160dt_default-drag}, the absolute error escalates to approximately 50\% after around 500 repetitions. However, by incorporating the custom control pulse, the absolute error remains under 20\%, even when subjected to nearly 1000 repetitions. Furthermore, it is noteworthy that employing a $17.8$ $ns$ pulse reduces the circuit latency to just about $17.78$ $\mu s$ after approximately 1000 repetitions. In comparison, the default Qiskit implementation, considering the same circuit duration, achieves word lengths of only approximately 250. However, with a word length of around 1000, the default implementation requires nearly $71.11$ $\mu s$, which is four times more than the custom implementation.

This result confirms the discussion in Section \ref{sec:fast-square-gate-performance}. For repetitive quantum circuits like those in QFA, calibrating fast and custom gatesets is essential. Square pulses are generally considered a significant source of leakage. However, in the case of this automaton, the consistent amplitude and lower values of square pulses are particularly interesting. If a Gaussian pulse were used instead, higher amplitude values would be necessary to maintain the same area-under-the-curve, and there would be considerable variations in the amplitude value, which could introduce amplitude control errors from the pulse-generating equipment and calibration errors due to inaccuracies. In contrast, the continuity of the same amplitude value in square pulses simplifies calibration and minimizes amplitude control errors, as there are no variations in the amplitude value.

\subsection{Summary}

Table \ref{tab:summary} presents a summary of the results obtained from comparing the default Qiskit gate implementation and the $17.8$ $ns$ square pulse. The table showcases the maximum word length supported by each implementation while considering predefined error thresholds of 10\% and 20\% in absolute error.

\begin{table}[h]
\caption{\label{tab:summary}Max word length supported by the different implementations considering an absolute error threshold of 10 and 20\% for Jakarta quantum computer.}
\begin{tabular}{ccccccc}
 & \textrm{threshold} & \textrm{$\mod$} & \textrm{Qiskit} & \textrm{Square} \\
\hline
& \multirow{2}{*}{\textrm{10\%}} & 0 & 66 & 539 \\
\textrm{max} && 3 & 58 & 795 \\
\textrm{length} & \multirow{2}{*}{20\%} & 0 & 132 & $>$1000 \\
&& 3 & 135 & $>$1000 \\
\end{tabular}
\end{table}

When applying a threshold of 20\%, the custom square pulse exhibited exceptional computational capabilities by supporting word lengths that surpassed 1000. In contrast, the default Qiskit implementation was constrained to a maximum word length of 135, exclusively for word lengths satisfying the condition 3 mod 11. When considering both word categories and the 10\% threshold, the automaton showcased very good performance, even with word lengths exceeding 500. This achievement was made possible because of the use of a simpler and faster control pulse, which distances the circuit latency from the decoherence times of the quantum computer. In comparison, the default Qiskit implementation was constrained to a word length of 66. Additionally, the results demonstrate significant enhancements in the maximum word length achievable for secure computations within the specified thresholds. Specifically, there were improvements of 12x and $>$7.4x for thresholds of 10\% and 20\%, respectively.

\section{Conclusions and Future Works}\label{sec:conclusion}

This study explored the implementation of a problem solved by means of a quantum finite automaton on a real quantum computer, considering the use of pulse-level programming for improving the performance in the computation of long strings. Default implementations had shown inefficient for executing long quantum automaton circuits. This inefficiency primarily arises because these gates are designed to be error-resistant and require minimal calibration. In the default approach, a sequence of two $71.11$ $ns$ DRAG pulses is employed to implement a $\sqrt{X}$ gate together with phase shifts. While this technique can implement every single-qubit gate, and therefore, reduce calibration requirements, it significantly extends circuit latency. In contrast, the proposed approach involved the calibration of a simple and fast square pulse lasting only $17.8$ $ns$ for the desired quantum gate. This calibration reduced circuit latency by a quarter.

With a fourfold reduction in gate duration, the automaton is now capable of processing strings with lengths approaching 1000 while maintaining an absolute error threshold of 20\%. In contrast, the default implementation was limited to word lengths slightly exceeding 130 under the same constraint. In comparison to other implementations found in the literature, our approach stands out as a 1-qubit implementation for the quantum automaton. Although there are some approaches with good results in the literature, they rely on artificial circuit optimization.

Future works in this field could leverage pulse-level programming to enhance the implementation of other Quantum Finite Automata (QFA) models, including pushdown automata. However, there remains a significant body of work to comprehensively grasp quantum automata and their language capabilities, along with their practical applications. In the context of implementation, pulse-level programming opens up a multitude of possibilities. While our implementation has demonstrated the superiority of simple short pulses, it is worth exploring alternative pulse-shaping techniques that might yield further advancements in quantum automata implementation.

\bmhead{Acknowledgements}

We express our sincere gratitude to Q-CTRL for their generous provision of notebooks, comprehensive learning materials, and practical insights into pulse-level programming. A special note of gratitude is extended to Dr. André Carvalho, the Head of Quantum Control Solutions at Q-CTRL, for his unwavering support in this endeavor. We also thank MSc. Evandro Chagas Ribeiro da Rosa for his valuable comments and contributions. This work was supported by Conselho Nacional de Desenvolvimento Científico e Tecnológico - CNPq through grant number 409673/2022-6.

\begin{appendices}

\section{Backend Configuration\label{apx:jobs-info}}

In this section, we provide Table \ref{tab:jobs-configuration} with information about the IBMQ jobs and backend configurations associated with all the demonstrations conducted on the Jakarta quantum computer and discussed in this paper.

\begin{table}[!h]
\caption{\label{tab:jobs-configuration}Jobs information.}
\begin{tabular}{ccccccc}
 & \textrm{$\mod$} & \textrm{date (UTC)} & \textrm{T1 ($\mu s$)} & \textrm{T2 ($\mu s$)} & \textrm{readout error} \\
\hline
\multirow{8}{*}{\textrm{Qiskit}}
 & \multirow{4}{*}{\textrm{0}} 
   & 08/15/23, 17:10 & 149.89 & 45.57 & 0.0166 \\
  && 08/16/23, 10:40 & 152.24 & 46.15 & 0.0155 \\
  && 08/16/23, 19:21 & 174.28 & 46.15 & 0.0155 \\
  && 08/17/23, 17:30 & 79.83  & 41.78 & 0.0197 \\
 & \multirow{4}{*}{\textrm{3}} 
   & 08/15/23, 17:10 & 149.89 & 45.57 & 0.0166 \\
  && 08/16/23, 10:40 & 152.24 & 46.15 & 0.0155 \\
  && 08/16/23, 19:21 & 174.28 & 46.15 & 0.0155 \\
  && 08/17/23, 20:40 & 138.63 & 41.78 & 0.0197 \\
\hline
\multirow{8}{*}{\textrm{Square}}
 & \multirow{4}{*}{\textrm{0}} 
   & 08/15/23, 19:20 & 100.76 & 45.57 & 0.0166 \\
  && 08/16/23, 10:40 & 152.24 & 46.15 & 0.0155 \\
  && 08/16/23, 19:21 & 174.28 & 46.15 & 0.0197 \\
  && 08/17/23, 20:40 & 138.63 & 41.78 & 0.0166 \\
 & \multirow{4}{*}{\textrm{3}} 
   & 08/15/23, 19:20 & 100.76 & 45.57 & 0.0166 \\
  && 08/16/23, 10:40 & 152.24 & 46.15 & 0.0155 \\
  && 08/16/23, 19:21 & 174.28 & 46.15 & 0.0155 \\
  && 08/17/23, 20:40 & 138.63 & 41.78 & 0.0197 \\
                                
\end{tabular}
\end{table}

\end{appendices}

\bibliography{sn-bibliography}


\begin{thebibliography}{49}
\ifx \bisbn   \undefined \def \bisbn  #1{ISBN #1}\fi
\ifx \binits  \undefined \def \binits#1{#1}\fi
\ifx \bauthor  \undefined \def \bauthor#1{#1}\fi
\ifx \batitle  \undefined \def \batitle#1{#1}\fi
\ifx \bjtitle  \undefined \def \bjtitle#1{#1}\fi
\ifx \bvolume  \undefined \def \bvolume#1{\textbf{#1}}\fi
\ifx \byear  \undefined \def \byear#1{#1}\fi
\ifx \bissue  \undefined \def \bissue#1{#1}\fi
\ifx \bfpage  \undefined \def \bfpage#1{#1}\fi
\ifx \blpage  \undefined \def \blpage #1{#1}\fi
\ifx \burl  \undefined \def \burl#1{\textsf{#1}}\fi
\ifx \doiurl  \undefined \def \doiurl#1{\url{https://doi.org/#1}}\fi
\ifx \betal  \undefined \def \betal{\textit{et al.}}\fi
\ifx \binstitute  \undefined \def \binstitute#1{#1}\fi
\ifx \binstitutionaled  \undefined \def \binstitutionaled#1{#1}\fi
\ifx \bctitle  \undefined \def \bctitle#1{#1}\fi
\ifx \beditor  \undefined \def \beditor#1{#1}\fi
\ifx \bpublisher  \undefined \def \bpublisher#1{#1}\fi
\ifx \bbtitle  \undefined \def \bbtitle#1{#1}\fi
\ifx \bedition  \undefined \def \bedition#1{#1}\fi
\ifx \bseriesno  \undefined \def \bseriesno#1{#1}\fi
\ifx \blocation  \undefined \def \blocation#1{#1}\fi
\ifx \bsertitle  \undefined \def \bsertitle#1{#1}\fi
\ifx \bsnm \undefined \def \bsnm#1{#1}\fi
\ifx \bsuffix \undefined \def \bsuffix#1{#1}\fi
\ifx \bparticle \undefined \def \bparticle#1{#1}\fi
\ifx \barticle \undefined \def \barticle#1{#1}\fi
\bibcommenthead
\ifx \bconfdate \undefined \def \bconfdate #1{#1}\fi
\ifx \botherref \undefined \def \botherref #1{#1}\fi
\ifx \url \undefined \def \url#1{\textsf{#1}}\fi
\ifx \bchapter \undefined \def \bchapter#1{#1}\fi
\ifx \bbook \undefined \def \bbook#1{#1}\fi
\ifx \bcomment \undefined \def \bcomment#1{#1}\fi
\ifx \oauthor \undefined \def \oauthor#1{#1}\fi
\ifx \citeauthoryear \undefined \def \citeauthoryear#1{#1}\fi
\ifx \endbibitem  \undefined \def \endbibitem {}\fi
\ifx \bconflocation  \undefined \def \bconflocation#1{#1}\fi
\ifx \arxivurl  \undefined \def \arxivurl#1{\textsf{#1}}\fi
\csname PreBibitemsHook\endcsname

\bibitem[\protect\citeauthoryear{Sipser}{2013}]{sipse2012introduction}
\begin{bbook}
\bauthor{\bsnm{Sipser}, \binits{M.}}:
\bbtitle{Introduction to the Theory of Computation},
\bedition{3}rd edn.
\bpublisher{Course Technology},
\blocation{Boston, MA}
(\byear{2013})
\end{bbook}
\endbibitem

\bibitem[\protect\citeauthoryear{Kondacs and Watrous}{1997}]{MM1QFA}
\begin{bchapter}
\bauthor{\bsnm{Kondacs}, \binits{A.}},
\bauthor{\bsnm{Watrous}, \binits{J.}}:
\bctitle{On the power of quantum finite state automata}.
In: \bbtitle{Proceedings 38th Annual Symposium on Foundations of Computer Science},
pp. \bfpage{66}--\blpage{75}
(\byear{1997}).
\doiurl{10.1109/SFCS.1997.646094}
\end{bchapter}
\endbibitem

\bibitem[\protect\citeauthoryear{Mereghetti et~al.}{2020}]{PhotonicQFA}
\begin{barticle}
\bauthor{\bsnm{Mereghetti}, \binits{C.}},
\bauthor{\bsnm{Palano}, \binits{B.}},
\bauthor{\bsnm{Cialdi}, \binits{S.}},
\bauthor{\bsnm{Vento}, \binits{V.}},
\bauthor{\bsnm{Paris}, \binits{M.G.A.}},
\bauthor{\bsnm{Olivares}, \binits{S.}}:
\batitle{Photonic realization of a quantum finite automaton}.
\bjtitle{Phys. Rev. Res.}
\bvolume{2},
\bfpage{013089}
(\byear{2020})
\doiurl{10.1103/PhysRevResearch.2.013089}
\end{barticle}
\endbibitem

\bibitem[\protect\citeauthoryear{Candeloro et~al.}{2021}]{EnhancedPhotonicQFA}
\begin{botherref}
\oauthor{\bsnm{Candeloro}, \binits{A.}},
\oauthor{\bsnm{Mereghetti}, \binits{C.}},
\oauthor{\bsnm{Palano}, \binits{B.}},
\oauthor{\bsnm{Cialdi}, \binits{S.}},
\oauthor{\bsnm{Paris}, \binits{M.G.A.}},
\oauthor{\bsnm{Olivares}, \binits{S.}}:
An enhanced photonic quantum finite automaton.
Applied Sciences
\textbf{11}(18)
(2021)
\doiurl{10.3390/app11188768}
\end{botherref}
\endbibitem

\bibitem[\protect\citeauthoryear{IBM}{2023}]{IBMQ}
\begin{botherref}
\oauthor{\bsnm{IBM}}:
IBM Quantum Computing.
\url{https://www.ibm.com/quantum-computing/}
(2023)
\end{botherref}
\endbibitem

\bibitem[\protect\citeauthoryear{Birkan et~al.}{2021}]{QFANoisy}
\begin{bchapter}
\bauthor{\bsnm{Birkan}, \binits{U.}},
\bauthor{\bsnm{Salehi}, \binits{{\"O}.}},
\bauthor{\bsnm{Olejar}, \binits{V.}},
\bauthor{\bsnm{Nurlu}, \binits{C.}},
\bauthor{\bsnm{Yakary{\i}lmaz}, \binits{A.}}:
\bctitle{Implementing quantum finite automata algorithms on noisy devices}.
In: \beditor{\bsnm{Paszynski}, \binits{M.}},
\beditor{\bsnm{Kranzlm{\"u}ller}, \binits{D.}},
\beditor{\bsnm{Krzhizhanovskaya}, \binits{V.V.}},
\beditor{\bsnm{Dongarra}, \binits{J.J.}},
\beditor{\bsnm{Sloot}, \binits{P.M.A.}} (eds.)
\bbtitle{Computational Science -- ICCS 2021},
pp. \bfpage{3}--\blpage{16}.
\bpublisher{Springer},
\blocation{Cham}
(\byear{2021})
\end{bchapter}
\endbibitem

\bibitem[\protect\citeauthoryear{Arute et~al.}{2019}]{QuantumSupremacy}
\begin{barticle}
\bauthor{\bsnm{Arute}, \binits{F.}},
\bauthor{\bsnm{Arya}, \binits{K.}},
\bauthor{\bsnm{Babbush}, \binits{R.}},
\bauthor{\bsnm{Bacon}, \binits{D.}},
\bauthor{\bsnm{Bardin}, \binits{J.C.}},
\bauthor{\bsnm{Barends}, \binits{R.}},
\bauthor{\bsnm{Biswas}, \binits{R.}},
\bauthor{\bsnm{Boixo}, \binits{S.}},
\bauthor{\bsnm{Brandao}, \binits{F.G.S.L.}},
\bauthor{\bsnm{Buell}, \binits{D.A.}}, \betal:
\batitle{Quantum supremacy using a programmable superconducting processor}.
\bjtitle{Nature}
\bvolume{574}(\bissue{7779}),
\bfpage{505}--\blpage{510}
(\byear{2019})
\doiurl{10.1038/s41586-019-1666-5}
\end{barticle}
\endbibitem

\bibitem[\protect\citeauthoryear{Wu et~al.}{2021}]{wu2021strong}
\begin{barticle}
\bauthor{\bsnm{Wu}, \binits{Y.}},
\bauthor{\bsnm{Bao}, \binits{W.-S.}},
\bauthor{\bsnm{Cao}, \binits{S.}},
\bauthor{\bsnm{Chen}, \binits{F.}},
\bauthor{\bsnm{Chen}, \binits{M.-C.}},
\bauthor{\bsnm{Chen}, \binits{X.}},
\bauthor{\bsnm{Chung}, \binits{T.-H.}},
\bauthor{\bsnm{Deng}, \binits{H.}},
\bauthor{\bsnm{Du}, \binits{Y.}},
\bauthor{\bsnm{Fan}, \binits{D.}}, \betal:
\batitle{Strong quantum computational advantage using a superconducting quantum processor}.
\bjtitle{Phys. Rev. Lett.}
\bvolume{127},
\bfpage{180501}
(\byear{2021})
\doiurl{10.1103/PhysRevLett.127.180501}
\end{barticle}
\endbibitem

\bibitem[\protect\citeauthoryear{Zhong et~al.}{2021}]{zhong2021phase}
\begin{barticle}
\bauthor{\bsnm{Zhong}, \binits{H.-S.}},
\bauthor{\bsnm{Deng}, \binits{Y.-H.}},
\bauthor{\bsnm{Qin}, \binits{J.}},
\bauthor{\bsnm{Wang}, \binits{H.}},
\bauthor{\bsnm{Chen}, \binits{M.-C.}},
\bauthor{\bsnm{Peng}, \binits{L.-C.}},
\bauthor{\bsnm{Luo}, \binits{Y.-H.}},
\bauthor{\bsnm{Wu}, \binits{D.}},
\bauthor{\bsnm{Gong}, \binits{S.-Q.}},
\bauthor{\bsnm{Su}, \binits{H.}}, \betal:
\batitle{Phase-programmable gaussian boson sampling using stimulated squeezed light}.
\bjtitle{Phys. Rev. Lett.}
\bvolume{127},
\bfpage{180502}
(\byear{2021})
\doiurl{10.1103/PhysRevLett.127.180502}
\end{barticle}
\endbibitem

\bibitem[\protect\citeauthoryear{Zhu et~al.}{2022}]{zhu2022quantum}
\begin{barticle}
\bauthor{\bsnm{Zhu}, \binits{Q.}},
\bauthor{\bsnm{Cao}, \binits{S.}},
\bauthor{\bsnm{Chen}, \binits{F.}},
\bauthor{\bsnm{Chen}, \binits{M.-C.}},
\bauthor{\bsnm{Chen}, \binits{X.}},
\bauthor{\bsnm{Chung}, \binits{T.-H.}},
\bauthor{\bsnm{Deng}, \binits{H.}},
\bauthor{\bsnm{Du}, \binits{Y.}},
\bauthor{\bsnm{Fan}, \binits{D.}},
\bauthor{\bsnm{Gong}, \binits{M.}}, \betal:
\batitle{Quantum computational advantage via 60-qubit 24-cycle random circuit sampling}.
\bjtitle{Science bulletin}
\bvolume{67}(\bissue{3}),
\bfpage{240}--\blpage{245}
(\byear{2022})
\end{barticle}
\endbibitem

\bibitem[\protect\citeauthoryear{Zhong et~al.}{2020}]{zhong2020quantum}
\begin{barticle}
\bauthor{\bsnm{Zhong}, \binits{H.-S.}},
\bauthor{\bsnm{Wang}, \binits{H.}},
\bauthor{\bsnm{Deng}, \binits{Y.-H.}},
\bauthor{\bsnm{Chen}, \binits{M.-C.}},
\bauthor{\bsnm{Peng}, \binits{L.-C.}},
\bauthor{\bsnm{Luo}, \binits{Y.-H.}},
\bauthor{\bsnm{Qin}, \binits{J.}},
\bauthor{\bsnm{Wu}, \binits{D.}},
\bauthor{\bsnm{Ding}, \binits{X.}},
\bauthor{\bsnm{Hu}, \binits{Y.}}, \betal:
\batitle{Quantum computational advantage using photons}.
\bjtitle{Science}
\bvolume{370}(\bissue{6523}),
\bfpage{1460}--\blpage{1463}
(\byear{2020})
\end{barticle}
\endbibitem

\bibitem[\protect\citeauthoryear{Madsen et~al.}{2022}]{madsen2022quantum}
\begin{barticle}
\bauthor{\bsnm{Madsen}, \binits{L.S.}},
\bauthor{\bsnm{Laudenbach}, \binits{F.}},
\bauthor{\bsnm{Askarani}, \binits{M.F.}},
\bauthor{\bsnm{Rortais}, \binits{F.}},
\bauthor{\bsnm{Vincent}, \binits{T.}},
\bauthor{\bsnm{Bulmer}, \binits{J.F.}},
\bauthor{\bsnm{Miatto}, \binits{F.M.}},
\bauthor{\bsnm{Neuhaus}, \binits{L.}},
\bauthor{\bsnm{Helt}, \binits{L.G.}},
\bauthor{\bsnm{Collins}, \binits{M.J.}}, \betal:
\batitle{Quantum computational advantage with a programmable photonic processor}.
\bjtitle{Nature}
\bvolume{606}(\bissue{7912}),
\bfpage{75}--\blpage{81}
(\byear{2022})
\end{barticle}
\endbibitem

\bibitem[\protect\citeauthoryear{Krantz et~al.}{2019}]{KrantzQuantumEngineersGuide}
\begin{botherref}
\oauthor{\bsnm{Krantz}, \binits{P.}},
\oauthor{\bsnm{Kjaergaard}, \binits{M.}},
\oauthor{\bsnm{Yan}, \binits{F.}},
\oauthor{\bsnm{Orlando}, \binits{T.P.}},
\oauthor{\bsnm{Gustavsson}, \binits{S.}},
\oauthor{\bsnm{Oliver}, \binits{W.D.}}:
A quantum engineer's guide to superconducting qubits.
Applied Physics Reviews
\textbf{6}
(2019)
\doiurl{10.1063/1.5089550}
\end{botherref}
\endbibitem

\bibitem[\protect\citeauthoryear{Maslov et~al.}{2018}]{OutlookQC}
\begin{barticle}
\bauthor{\bsnm{Maslov}, \binits{D.}},
\bauthor{\bsnm{Nam}, \binits{Y.}},
\bauthor{\bsnm{Kim}, \binits{J.}}:
\batitle{An outlook for quantum computing}.
\bjtitle{Point of View}
\bvolume{107}(\bissue{1}),
\bfpage{5}--\blpage{10}
(\byear{2018})
\doiurl{10.1109/JPROC.2018.2884353}
\end{barticle}
\endbibitem

\bibitem[\protect\citeauthoryear{Q-CTRL}{2023}]{QCTRL_QuantumControl}
\begin{botherref}
\oauthor{\bsnm{Q-CTRL}}:
What Is Quantum Control?
\end{botherref}
\endbibitem

\bibitem[\protect\citeauthoryear{Qiskit}{2023}]{QiskitPulseDoc}
\begin{botherref}
\oauthor{\bsnm{Qiskit}, \binits{I.}}:
Qiskit Pulse.
IBMQ
(2023)
\end{botherref}
\endbibitem

\bibitem[\protect\citeauthoryear{Alexander et~al.}{2020}]{QiskitPulse}
\begin{botherref}
\oauthor{\bsnm{Alexander}, \binits{T.}},
\oauthor{\bsnm{Kanazawa}, \binits{N.}},
\oauthor{\bsnm{Egger}, \binits{D.J.}},
\oauthor{\bsnm{Capelluto}, \binits{L.}},
\oauthor{\bsnm{Wood}, \binits{C.J.}},
\oauthor{\bsnm{Javadi-Abhari}, \binits{A.}},
\oauthor{\bsnm{McKay}, \binits{D.C.}}:
Qiskit pulse: Programming quantum computers through the cloud with pulses.
Quantum Science and Technology
\textbf{5}
(2020)
\doiurl{10.1088/2058-9565/aba404}
\end{botherref}
\endbibitem

\bibitem[\protect\citeauthoryear{Carvalho et~al.}{2021}]{ErrorRobustAndre}
\begin{botherref}
\oauthor{\bsnm{Carvalho}, \binits{A.R.R.}},
\oauthor{\bsnm{Bal}, \binits{H.}},
\oauthor{\bsnm{Biercuk}, \binits{M.J.}},
\oauthor{\bsnm{Hush}, \binits{M.R.}},
\oauthor{\bsnm{Thomsen}, \binits{F.}}:
Error-robust quantum logic optimization using a cloud quantum computing interface.
Physical Review Applied
\textbf{15}(6)
(2021)
\end{botherref}
\endbibitem

\bibitem[\protect\citeauthoryear{Baum et~al.}{2021}]{DRL_QCTRL}
\begin{botherref}
\oauthor{\bsnm{Baum}, \binits{Y.}},
\oauthor{\bsnm{Amico}, \binits{M.}},
\oauthor{\bsnm{Howell}, \binits{S.}},
\oauthor{\bsnm{Hush}, \binits{M.}},
\oauthor{\bsnm{Liuzzi}, \binits{M.}},
\oauthor{\bsnm{Mundada}, \binits{P.}},
\oauthor{\bsnm{Merkh}, \binits{T.}},
\oauthor{\bsnm{Carvalho}, \binits{A.R.R.}},
\oauthor{\bsnm{Biercuk}, \binits{M.J.}}:
Experimental deep reinforcement learning for error-robust gate-set design on a superconducting quantum computer.
PRX Quantum
\textbf{2}(4)
(2021)
\end{botherref}
\endbibitem

\bibitem[\protect\citeauthoryear{McEwen et~al.}{2021}]{Leakage}
\begin{botherref}
\oauthor{\bsnm{McEwen}, \binits{M.}},
\oauthor{\bsnm{Kafri}, \binits{D.}},
\oauthor{\bsnm{Chen}, \binits{Z.}},
\oauthor{\bsnm{Atalaya}, \binits{J.}},
\oauthor{\bsnm{Satzinger}, \binits{K.J.}},
\oauthor{\bsnm{Quintana}, \binits{C.}},
\oauthor{\bsnm{Klimov}, \binits{P.V.}},
\oauthor{\bsnm{Sank}, \binits{D.}},
\oauthor{\bsnm{Gidney}, \binits{C.}},
\oauthor{\bsnm{Fowler}, \binits{A.G.}}, et al.:
Removing leakage-induced correlated errors in superconducting quantum error correction.
Nature Communications
\textbf{12}
(2021)
\end{botherref}
\endbibitem

\bibitem[\protect\citeauthoryear{Gertler et~al.}{2021}]{Bosonic}
\begin{barticle}
\bauthor{\bsnm{Gertler}, \binits{J.M.}},
\bauthor{\bsnm{Baker}, \binits{B.}},
\bauthor{\bsnm{Li}, \binits{J.}},
\bauthor{\bsnm{Shirol}, \binits{S.}},
\bauthor{\bsnm{Koch}, \binits{J.}},
\bauthor{\bsnm{Wang}, \binits{C.}}:
\batitle{Protecting a bosonic qubit with autonomous quantum error correction}.
\bjtitle{Nature}
\bvolume{590},
\bfpage{243}--\blpage{248}
(\byear{2021})
\doiurl{10.1038/s41586-021-03257-0}
\end{barticle}
\endbibitem

\bibitem[\protect\citeauthoryear{Fu et~al.}{2019}]{microarchitecture}
\begin{barticle}
\bauthor{\bsnm{Fu}, \binits{X.}},
\bauthor{\bsnm{Lao}, \binits{L.}},
\bauthor{\bsnm{Bertels}, \binits{K.}},
\bauthor{\bsnm{Almudever}, \binits{C.G.}}:
\batitle{A control microarchitecture for fault-tolerant quantum computing}.
\bjtitle{Microprocessors and Microsystems}
\bvolume{70},
\bfpage{21}--\blpage{30}
(\byear{2019})
\doiurl{10.1016/j.micpro.2019.06.011}
\end{barticle}
\endbibitem

\bibitem[\protect\citeauthoryear{Palittapongarnpim et~al.}{2017}]{Learning}
\begin{barticle}
\bauthor{\bsnm{Palittapongarnpim}, \binits{P.}},
\bauthor{\bsnm{Wittek}, \binits{P.}},
\bauthor{\bsnm{Zahedinejad}, \binits{E.}},
\bauthor{\bsnm{Vedaie}, \binits{S.}},
\bauthor{\bsnm{Sanders}, \binits{B.C.}}:
\batitle{Learning in quantum control: High-dimensional global optimization for noisy quantum dynamics}.
\bjtitle{Neurocomputing}
\bvolume{268},
\bfpage{116}--\blpage{126}
(\byear{2017})
\doiurl{10.1016/j.neucom.2016.12.087} .
\bcomment{Advances in artificial neural networks, machine learning and computational intelligence}
\end{barticle}
\endbibitem

\bibitem[\protect\citeauthoryear{Propson et~al.}{2022}]{Propson2022}
\begin{barticle}
\bauthor{\bsnm{Propson}, \binits{T.}},
\bauthor{\bsnm{Jackson}, \binits{B.E.}},
\bauthor{\bsnm{Koch}, \binits{J.}},
\bauthor{\bsnm{Manchester}, \binits{Z.}},
\bauthor{\bsnm{Schuster}, \binits{D.I.}}:
\batitle{Robust quantum optimal control with trajectory optimization}.
\bjtitle{Phys. Rev. Appl.}
\bvolume{17},
\bfpage{014036}
(\byear{2022})
\doiurl{10.1103/PhysRevApplied.17.014036}
\end{barticle}
\endbibitem

\bibitem[\protect\citeauthoryear{Mundada et~al.}{2023}]{Mundada2023}
\begin{barticle}
\bauthor{\bsnm{Mundada}, \binits{P.S.}},
\bauthor{\bsnm{Barbosa}, \binits{A.}},
\bauthor{\bsnm{Maity}, \binits{S.}},
\bauthor{\bsnm{Wang}, \binits{Y.}},
\bauthor{\bsnm{Merkh}, \binits{T.}},
\bauthor{\bsnm{Stace}, \binits{T.M.}},
\bauthor{\bsnm{Nielson}, \binits{F.}},
\bauthor{\bsnm{Carvalho}, \binits{A.R.R.}},
\bauthor{\bsnm{Hush}, \binits{M.}},
\bauthor{\bsnm{Biercuk}, \binits{M.J.}},
\bauthor{\bsnm{Baum}, \binits{Y.}}:
\batitle{Experimental benchmarking of an automated deterministic error-suppression workflow for quantum algorithms}.
\bjtitle{Phys. Rev. Appl.}
\bvolume{20},
\bfpage{024034}
(\byear{2023})
\doiurl{10.1103/PhysRevApplied.20.024034}
\end{barticle}
\endbibitem

\bibitem[\protect\citeauthoryear{Q-CTRL}{2023}]{QCTRL_site}
\begin{botherref}
\oauthor{\bsnm{Q-CTRL}}:
Homepage.
\url{https://q-ctrl.com/}
(2023)
\end{botherref}
\endbibitem

\bibitem[\protect\citeauthoryear{Nautrup et~al.}{2019}]{Nautrup2019optimizingquantum}
\begin{barticle}
\bauthor{\bsnm{Nautrup}, \binits{H.P.}},
\bauthor{\bsnm{Delfosse}, \binits{N.}},
\bauthor{\bsnm{Dunjko}, \binits{V.}},
\bauthor{\bsnm{Briegel}, \binits{H.J.}},
\bauthor{\bsnm{Friis}, \binits{N.}}:
\batitle{Optimizing {Q}uantum {E}rror {C}orrection {C}odes with {R}einforcement {L}earning}.
\bjtitle{{Quantum}}
\bvolume{3},
\bfpage{215}
(\byear{2019})
\doiurl{10.22331/q-2019-12-16-215}
\end{barticle}
\endbibitem

\bibitem[\protect\citeauthoryear{Moore and Crutchfield}{2000}]{moore2000quantum}
\begin{barticle}
\bauthor{\bsnm{Moore}, \binits{C.}},
\bauthor{\bsnm{Crutchfield}, \binits{J.P.}}:
\batitle{Quantum automata and quantum grammars}.
\bjtitle{Theoretical Computer Science}
\bvolume{237}(\bissue{1-2}),
\bfpage{275}--\blpage{306}
(\byear{2000})
\end{barticle}
\endbibitem

\bibitem[\protect\citeauthoryear{IBM}{2023}]{IBMQE}
\begin{botherref}
\oauthor{\bsnm{IBM}}:
IBM Q Experience.
\url{https://www.ibm.com/quantum-computing/technology/experience/}
(2023)
\end{botherref}
\endbibitem

\bibitem[\protect\citeauthoryear{Ambainis and Watrous}{2002}]{2CQFA}
\begin{botherref}
\oauthor{\bsnm{Ambainis}, \binits{A.}},
\oauthor{\bsnm{Watrous}, \binits{J.}}:
Two-way finite automata with quantum and classical states.
Theoretical Computer Science
\textbf{287}
(2002)
\doiurl{10.1016/S0304-3975(02)00138-X}
\end{botherref}
\endbibitem

\bibitem[\protect\citeauthoryear{Amano and Iwama}{1999}]{15QFA}
\begin{bchapter}
\bauthor{\bsnm{Amano}, \binits{M.}},
\bauthor{\bsnm{Iwama}, \binits{K.}}:
\bctitle{Undecidability on quantum finite automata}.
In: \bbtitle{Proceedings of the Thirty-First Annual ACM Symposium on Theory of Computing}.
\bsertitle{STOC '99},
pp. \bfpage{368}--\blpage{375}.
\bpublisher{Association for Computing Machinery},
\blocation{New York, NY, USA}
(\byear{1999}).
\doiurl{10.1145/301250.301344} .
\burl{https://doi.org/10.1145/301250.301344}
\end{bchapter}
\endbibitem

\bibitem[\protect\citeauthoryear{Say and Yakaryılmaz}{2014}]{say2014quantum}
\begin{bbook}
\bauthor{\bsnm{Say}, \binits{A.C.C.}},
\bauthor{\bsnm{Yakaryılmaz}, \binits{A.}}:
\bbtitle{Quantum Finite Automata: A Modern Introduction},
pp. \bfpage{208}--\blpage{222}.
\bpublisher{Springer},
\blocation{Cham}
(\byear{2014}).
\doiurl{10.1007/978-3-319-13350-8_16}
\end{bbook}
\endbibitem

\bibitem[\protect\citeauthoryear{Ambainis and Freivalds}{1998}]{ambainis19981}
\begin{bchapter}
\bauthor{\bsnm{Ambainis}, \binits{A.}},
\bauthor{\bsnm{Freivalds}, \binits{R.}}:
\bctitle{1-way quantum finite automata: strengths, weaknesses and generalizations}.
In: \bbtitle{Proceedings 39th Annual Symposium on Foundations of Computer Science (Cat. No. 98CB36280)},
pp. \bfpage{332}--\blpage{341}
(\byear{1998}).
\bcomment{IEEE}
\end{bchapter}
\endbibitem

\bibitem[\protect\citeauthoryear{Ballon}{2022}]{SuperconductingPennyLane}
\begin{botherref}
\oauthor{\bsnm{Ballon}, \binits{A.}}:
Quantum Computing with Superconducting Qubits.
\url{https://pennylane.ai/qml/demos/tutorial_sc_qubits.html}
(2022)
\end{botherref}
\endbibitem

\bibitem[\protect\citeauthoryear{IBM}{2022}]{CalibratingQiskit}
\begin{botherref}
\oauthor{\bsnm{IBM}}:
Calibrating Qubits with Qiski Pulse.
\url{https://qiskit.org/textbook/ch-quantum-hardware/calibrating-qubits-pulse.html}
(2022)
\end{botherref}
\endbibitem

\bibitem[\protect\citeauthoryear{Jo et~al.}{2019}]{LeakageFastPulse}
\begin{botherref}
\oauthor{\bsnm{Jo}, \binits{H.}},
\oauthor{\bsnm{Song}, \binits{Y.}},
\oauthor{\bsnm{Ahn}, \binits{J.}}:
Leakage suppression by ultrafast composite pulses.
Optics Express
\textbf{27}
(2019)
\doiurl{10.1364/OE.27.003944}
\end{botherref}
\endbibitem

\bibitem[\protect\citeauthoryear{Hui}{2018}]{QC_ImplementationIssues}
\begin{botherref}
\oauthor{\bsnm{Hui}, \binits{J.}}:
QC - Quantum programming and implementation issues.
Medium
(2018)
\end{botherref}
\endbibitem

\bibitem[\protect\citeauthoryear{Chen et~al.}{2016}]{Chen2016leakage}
\begin{botherref}
\oauthor{\bsnm{Chen}, \binits{Z.}},
\oauthor{\bsnm{Kelly}, \binits{J.}},
\oauthor{\bsnm{Quintana}, \binits{C.}},
\oauthor{\bsnm{Barends}, \binits{R.}},
\oauthor{\bsnm{Campbell}, \binits{B.}},
\oauthor{\bsnm{Chen}, \binits{Y.}},
\oauthor{\bsnm{Chiaro}, \binits{B.}},
\oauthor{\bsnm{Dunsworth}, \binits{A.}},
\oauthor{\bsnm{Fowler}, \binits{A.G.}},
\oauthor{\bsnm{Lucero}, \binits{E.}}, et al.:
Measuring and suppressing quantum state leakage in a superconducting qubit.
Physical Review Letters
\textbf{116}
(2016)
\doiurl{10.1103/PhysRevLett.116.020501}
\end{botherref}
\endbibitem

\bibitem[\protect\citeauthoryear{Sarovar et~al.}{2020}]{Sarovar2020detectingcrosstalk}
\begin{barticle}
\bauthor{\bsnm{Sarovar}, \binits{M.}},
\bauthor{\bsnm{Proctor}, \binits{T.}},
\bauthor{\bsnm{Rudinger}, \binits{K.}},
\bauthor{\bsnm{Young}, \binits{K.}},
\bauthor{\bsnm{Nielsen}, \binits{E.}},
\bauthor{\bsnm{Blume-Kohout}, \binits{R.}}:
\batitle{Detecting crosstalk errors in quantum information processors}.
\bjtitle{{Quantum}}
\bvolume{4},
\bfpage{321}
(\byear{2020})
\doiurl{10.22331/q-2020-09-11-321}
\end{barticle}
\endbibitem

\bibitem[\protect\citeauthoryear{Funcke et~al.}{2022}]{Funcke2022}
\begin{barticle}
\bauthor{\bsnm{Funcke}, \binits{L.}},
\bauthor{\bsnm{Hartung}, \binits{T.}},
\bauthor{\bsnm{Jansen}, \binits{K.}},
\bauthor{\bsnm{K\"uhn}, \binits{S.}},
\bauthor{\bsnm{Stornati}, \binits{P.}},
\bauthor{\bsnm{Wang}, \binits{X.}}:
\batitle{Measurement error mitigation in quantum computers through classical bit-flip correction}.
\bjtitle{Phys. Rev. A}
\bvolume{105},
\bfpage{062404}
(\byear{2022})
\doiurl{10.1103/PhysRevA.105.062404}
\end{barticle}
\endbibitem

\bibitem[\protect\citeauthoryear{Qiskit}{2023a}]{QiskitDocSystemInfo}
\begin{botherref}
\oauthor{\bsnm{Qiskit}, \binits{I.}}:
Gathering System Information.
IBMQ
(2023)
\end{botherref}
\endbibitem

\bibitem[\protect\citeauthoryear{Qiskit}{2023b}]{QiskitDocBuildSchedules}
\begin{botherref}
\oauthor{\bsnm{Qiskit}, \binits{I.}}:
Building Pulse Schedules.
IBMQ
(2023)
\end{botherref}
\endbibitem

\bibitem[\protect\citeauthoryear{Qiskit}{2023c}]{QiskitDocDRAG042}
\begin{botherref}
\oauthor{\bsnm{Qiskit}, \binits{I.}}:
Qiskit Pulse 0.42 - DRAG Documentation.
IBMQ
(2023)
\end{botherref}
\endbibitem

\bibitem[\protect\citeauthoryear{Qiskit}{}]{QiskitPassManager}
\begin{botherref}
\oauthor{\bsnm{Qiskit}}:
Transpiler Passes and Pass Manager.
\url{https://qiskit.org/documentation/tutorials/circuits_advanced/04_transpiler_passes_and_passmanager.html}
\end{botherref}
\endbibitem

\bibitem[\protect\citeauthoryear{McKay et~al.}{2017}]{McKay2017}
\begin{barticle}
\bauthor{\bsnm{McKay}, \binits{D.C.}},
\bauthor{\bsnm{Wood}, \binits{C.J.}},
\bauthor{\bsnm{Sheldon}, \binits{S.}},
\bauthor{\bsnm{Chow}, \binits{J.M.}},
\bauthor{\bsnm{Gambetta}, \binits{J.M.}}:
\batitle{Efficient $z$ gates for quantum computing}.
\bjtitle{Phys. Rev. A}
\bvolume{96},
\bfpage{022330}
(\byear{2017})
\doiurl{10.1103/PhysRevA.96.022330}
\end{barticle}
\endbibitem

\bibitem[\protect\citeauthoryear{Motzoi et~al.}{2009}]{Gambetta2009}
\begin{barticle}
\bauthor{\bsnm{Motzoi}, \binits{F.}},
\bauthor{\bsnm{Gambetta}, \binits{J.M.}},
\bauthor{\bsnm{Rebentrost}, \binits{P.}},
\bauthor{\bsnm{Wilhelm}, \binits{F.K.}}:
\batitle{Simple pulses for elimination of leakage in weakly nonlinear qubits}.
\bjtitle{Phys. Rev. Lett.}
\bvolume{103},
\bfpage{110501}
(\byear{2009})
\doiurl{10.1103/PhysRevLett.103.110501}
\end{barticle}
\endbibitem

\bibitem[\protect\citeauthoryear{Gambetta et~al.}{2011}]{Gambetta2011}
\begin{barticle}
\bauthor{\bsnm{Gambetta}, \binits{J.M.}},
\bauthor{\bsnm{Motzoi}, \binits{F.}},
\bauthor{\bsnm{Merkel}, \binits{S.T.}},
\bauthor{\bsnm{Wilhelm}, \binits{F.K.}}:
\batitle{Analytic control methods for high-fidelity unitary operations in a weakly nonlinear oscillator}.
\bjtitle{Phys. Rev. A}
\bvolume{83},
\bfpage{012308}
(\byear{2011})
\doiurl{10.1103/PhysRevA.83.012308}
\end{barticle}
\endbibitem

\bibitem[\protect\citeauthoryear{Q-CTRL}{2023a}]{QCTRLCalibration}
\begin{botherref}
\oauthor{\bsnm{Q-CTRL}}:
How to Automate Calibration of Control Hardware.
\url{https://docs.q-ctrl.com/boulder-opal/user-guides/how-to-automate-calibration-of-control-hardware}
(2023)
\end{botherref}
\endbibitem

\bibitem[\protect\citeauthoryear{Q-CTRL}{2023b}]{QCTRLDesignGates}
\begin{botherref}
\oauthor{\bsnm{Q-CTRL}}:
Designing Noise-Robust Single-Qubit Gates for IBM Qiskit.
\url{https://docs.q-ctrl.com/boulder-opal/application-notes/designing-noise-robust-single-qubit-gates-for-ibm-qiskit}
(2023)
\end{botherref}
\endbibitem

\end{thebibliography}

\end{document}